\documentclass[twocolumn,tighten,astrosymb,trackchanges]{aastex631}
\newcommand{\LBNL}{\affiliation{Lawrence Berkeley National Laboratory, 1 Cyclotron Road, Berkeley, CA 94720, USA}}

\newcommand{\LJMU}{\affiliation{Astrophysics Research Institute, Liverpool John Moores University, 146 Brownlow Hill, Liverpool L3 5RF, UK}}

\newcommand{\DIRAC}{\affiliation{DIRAC Institute, Department of Astronomy, University of Washington, 3910 15th Avenue NE, Seattle, WA 98195, USA}}

\newcommand{\HUBerlin}{\affiliation{Institut f\"ur Physik, Humboldt-Universit\"at zu Berlin, Newtonstr. 15, 12489 Berlin, Germany}}

\newcommand{\UMDAstro}{\affiliation{Department of Astronomy, University of Maryland, College Park, MD 20742, USA}}
\newcommand{\OKC}{\affiliation{Department of Physics, Oskar Klein Centre, Stockholm University, SE-106 91, Stockholm, Sweden}}

\newcommand{\Cornell}{\affiliation{Department of Astronomy, Cornell University, Ithaca, NY 14853, USA}}

\newcommand{\KIPAC}{\affiliation{Kavli Institute for Particle Astrophysics and Cosmology, Stanford, CA 94305, USA}}

\newcommand{\OKCAstro}{\affiliation{Department of Astronomy, Oskar Klein Center, Stockholm University, SE-106 91 Stockholm, Sweden}}

\newcommand{\IoACam}{\affiliation{Institute of Astronomy and Kavli Institute for Cosmology, University of Cambridge, Madingley Road, Cambridge, CB3 0HA, UK}}

\newcommand{\Caltech}{\affiliation{Cahill Center for Astrophysics, California Institute of Technology, 1200 E. California Boulevard, Pasadena, CA 91125, USA}}

\newcommand{\CaltechPhys}{\affiliation{Division of Physics, Mathematics and Astronomy, California Institute of Technology, Pasadena, CA 91125, USA}}

\newcommand{\GSFC}{\affiliation{Astrophysics Science Division, NASA Goddard Space Flight Center, Mail Code 661, Greenbelt, MD 20771, USA}}
\newcommand{\UMD}{\affiliation{Joint Space-Science Institute, University of Maryland, College Park, MD 20742, USA}}
\newcommand{\UCB}{\affiliation{Department of Astronomy, University of California, Berkeley, CA 94720-3411, USA}}

\newcommand{\IPAC}{\affiliation{IPAC, California Institute of Technology, 1200 E. California Blvd, Pasadena, CA 91125, USA}}

\newcommand{\NAOC}{\affiliation{National Astronomical Observatories, Chinese Academy of Sciences, A20 Datun Road, Chaoyang District, Beijing, 100101, P.~R.~China}}

\newcommand{\MPIA}{\affiliation{Max-Planck-Institut f\"ur Astrophysik, Karl-Schwarzschild-Stra\ss{}e 1, D-85748 Garching, Germany}}

\newcommand{\Birmingham}{\affiliation{Birmingham Institute for Gravitational Wave Astronomy and School of Physics and Astronomy, University of Birmingham, Birmingham B15 2TT, UK}}

\newcommand{\UNC}{\affiliation{Department of Physics and Astronomy, University of North Carolina, 120 East Cameron Avenue, Chapel Hill, NC 27599, USA}}

\newcommand{\CMU}{\affiliation{Department of Physics, Carnegie Mellon University, 5000 Forbes Avenue, Pittsburgh, PA 15213-3815, USA}}

\newcommand{\IfA}{\affiliation{Institute for Astronomy, University of Hawai`i, 2680 Woodlawn Drive, Honolulu, HI 96822-1839, USA}}

\newcommand{\CIERA}{\affiliation{Center for Interdisciplinary Exploration and Research in Astrophysics (CIERA), Northwestern University, 1800 Sherman Ave., Evanston, IL 60201, USA}}

\newcommand{\SkAI}{\affiliation{NSF-Simons AI Institute for the Sky (SkAI), 172 E. Chestnut St., Chicago, IL 60611, USA}}

\newcommand{\ICE}{\affiliation{Institute of Space Sciences (ICE, CSIC), Campus UAB, Carrer de Can Magrans, s/n, E-08193 Barcelona, Spain}}
\newcommand{\IEEC}{\affiliation{Institut d’Estudis Espacials de Catalunya (IEEC), E-08034 Barcelona, Spain}}

\newcommand{\ICG}{\affiliation{Institute of Cosmology and Gravitation, University of Portsmouth, Dennis Sciama Building, Burnaby Road, Portsmouth PO1 3FX, UK}}

\newcommand{\LPNHE}{\affiliation{Sorbonne Universit\'{e}, CNRS/IN2P3, Laboratoire de Physique Nucl\'{e}aire et de Hautes Energies (LPNHE), FR-75005 Paris, France}}

\newcommand{\UVa}{\affiliation{Department of Astronomy, University of Virginia, 530 McCormick Rd, Charlottesville, VA 22904, USA}}
\newcommand{\CaltechOO}{\affiliation{Caltech Optical Observatories, California Institute of Technology, Pasadena, CA 91125, USA}}

\newcommand{\mcwilliams}{\affiliation{McWilliams Center for Cosmology and Astrophysics, Department of Physics,     Carnegie Mellon University, 5000 Forbes Avenue, Pittsburgh, PA 15213, USA}}
\newcommand{\LMU}{\affiliation{University Observatory, Faculty of Physics, Ludwig-Maximilians-Universität, Scheinerstr. 1, 81679 Munich, Germany}}
\newcommand{\ORIGINS}{\affiliation{Excellence Cluster ORIGINS, Boltzmannstr. 2, 85748 Garching, Germany}}

\newcommand{\Lyon}{\affiliation{Universite Claude Bernard Lyon 1, CNRS, IP2I Lyon / IN2P3, IMR 5822, F-69622 Villeurbanne, France}}
\newcommand{\UIUC}{\affiliation{Department of Astronomy, University of Illinois Urbana-Champaign, 1002 West Green Street, Urbana, IL 61801, USA}}

\newcommand{\Lancaster}{\affiliation{Department of Physics, Lancaster University, Lancs LA1 4YB, UK}}

\newcommand{\Minnesota}{\affiliation{School of Physics and Astronomy, University of Minnesota, Minneapolis, Minnesota 55455, USA}}

\newcommand{\OSU}{\affiliation{Center for Cosmology and Astroparticle Physics, The Ohio State University, 191 West Woodruff Ave, Columbus, OH 43215, USA}}

\newcommand{\ROCH}{\affiliation{Department of Physics \& Astronomy, University of Rochester, 206 Bausch and Lomb Hall, P.O. Box 270171, Rochester, NY 14627-0171, USA}}

\newcommand{\BU}{\affiliation{Department of Physics, Boston University, 590 Commonwealth Avenue, Boston, MA 02215, USA}}

\newcommand{\UniMi}{\affiliation{Dipartimento di Fisica ``Aldo Pontremoli'', Universit\`a degli Studi di Milano, Via Celoria 16, I-20133 Milano, Italy}}

\newcommand{\INAFBrera}{\affiliation{INAF-Osservatorio Astronomico di Brera, Via Brera 28, 20122 Milano, Italy}}

\newcommand{\UCL}{\affiliation{Department of Physics \& Astronomy, University College London, Gower Street, London, WC1E 6BT, UK}}

\newcommand{\UNAMIF}{\affiliation{Instituto de F\'isica, Universidad Nacional Aut\'onoma de M\'exico, Circuito de la Investigaci\'on Cient\'ifica, Ciudad Universitaria, Cd. de M\'exico C.~P.~04510, M\'exico}}

\newcommand{\SDSU}{\affiliation{Department of Astronomy, San Diego State University, 5500 Campanile Drive, San Diego, CA 92182, USA}}

\newcommand{\NOIRLab}{\affiliation{NSF NOIRLab, 950 N. Cherry Ave., Tucson, AZ 85719, USA}}

\newcommand{\SMU}{\affiliation{Department of Physics, Southern Methodist University, 3215 Daniel Avenue, Dallas, TX 75275, USA}}

\newcommand{\Fermilab}{\affiliation{Fermi National Accelerator Laboratory, PO Box 500, Batavia, IL 60510, USA}}

\newcommand{\ICREA}{\affiliation{Instituci\'o Catalana de Recerca i Estudis Avan\c{c}ats, Passeig de Llu\'is Companys, 23, 08010 Barcelona, Spain}}

\newcommand{\IFAEBarcelona}{\affiliation{Institut de F\'isica d'Altes Energies (IFAE), The Barcelona Institute of Science and Technology, Edifici Cn, Campus UAB, 08193, Bellaterra (Barcelona), Spain}}

\newcommand{\DeFBarcelona}{\affiliation{Departament de F\'{i}sica, Serra H\'{u}nter, Universitat Aut\`{o}noma de Barcelona, 08193 Bellaterra (Barcelona), Spain}}

\newcommand{\UniandesPhys}{\affiliation{Departamento de F\'isica, Universidad de los Andes, Cra. 1 No. 18A-10, Edificio Ip, CP 111711, Bogot\'a, Colombia}}

\newcommand{\UniandesObs}{\affiliation{Observatorio Astron\'omico, Universidad de los Andes, Cra. 1 No. 18A-10, Edificio H, CP 111711, Bogot\'a, Colombia}}

\newcommand{\UMichPhys}{\affiliation{Department of Physics, University of Michigan, 450 Church Street, Ann Arbor, MI 48109, USA}}

\newcommand{\UMich}{\affiliation{University of Michigan, 500 S. State Street, Ann Arbor, MI 48109, USA}}

\newcommand{\UTDallas}{\affiliation{Department of Physics, The University of Texas at Dallas, 800 W. Campbell Rd., Richardson, TX 75080, USA}}

\newcommand{\IRFU}{\affiliation{IRFU, CEA, Universit\'{e} Paris-Saclay, F-91191 Gif-sur-Yvette, France}}

\newcommand{\CSIC}{\affiliation{Instituto de Astrof\'{i}sica de Andaluc\'{i}a (CSIC), Glorieta de la Astronom\'{i}a, s/n, E-18008 Granada, Spain}}

\newcommand{\UCIrvine}{\affiliation{Department of Physics and Astronomy, University of California, Irvine, 92697, USA}}

\newcommand{\Siena}{\affiliation{Department of Physics and Astronomy, Siena University, 515 Loudon Road, Loudonville, NY 12211, USA}}

\newcommand{\WaterlooPhysAstro}{\affiliation{Department of Physics and Astronomy, University of Waterloo, 200 University Ave W, Waterloo, ON N2L 3G1, Canada}}

\newcommand{\Perimeter}{\affiliation{Perimeter Institute for Theoretical Physics, 31 Caroline St. North, Waterloo, ON N2L 2Y5, Canada}}

\newcommand{\WaterlooCfA}{\affiliation{Waterloo Centre for Astrophysics, University of Waterloo, 200 University Ave W, Waterloo, ON N2L 3G1, Canada}}

\newcommand{\UCBSSL}{\affiliation{Space Sciences Laboratory, University of California, Berkeley, 7 Gauss Way, Berkeley, CA 94720, USA}}

\newcommand{\UCBcampus}{\affiliation{University of California, Berkeley, 110 Sproul Hall \#5800, Berkeley, CA 94720, USA}}

\newcommand{\UPCPhysEEBE}{\affiliation{Departament de F\'isica, EEBE, Universitat Polit\`ecnica de Catalunya, c/Eduard Maristany 10, 08930 Barcelona, Spain}}

\newcommand{\Sejong}{\affiliation{Department of Physics and Astronomy, Sejong University, 209 Neungdong-ro, Gwangjin-gu, Seoul 05006, Republic of Korea}}

\newcommand{\CIEMAT}{\affiliation{CIEMAT, Avenida Complutense 40, E-28040 Madrid, Spain}}

\newcommand{\KASI}{\affiliation{Korea Astronomy and Space Science Institute, 776, Daedeokdae-ro, Yuseong-gu, Daejeon 34055, Republic of Korea}}

\newcommand{\USTKorea}{\affiliation{University of Science and Technology, 217 Gajeong-ro, Yuseong-gu, Daejeon 34113, Republic of Korea}}

\usepackage{CJK}
\usepackage{multirow}
\newcommand\kms{~km~s$^{-1}$}

\let\tablenum\relax
\usepackage{siunitx}
\usepackage{appendix}
\usepackage{amsmath,amsthm,amsfonts,amssymb,amscd}
\usepackage{booktabs}
\usepackage{tablefootnote}
\usepackage{apjfonts}
\usepackage{lipsum}
\usepackage{changepage}
\usepackage{xcolor}
\usepackage[normalem]{ulem}
\usepackage{amssymb}

\let\ts=\thinspace
\newcommand{\one}{\ts {\sc i}}
\newcommand{\two}{\ts {\sc ii}}
\newcommand{\three}{\ts {\sc iii}}
\newcommand{\four}{\ts {\sc iv}}
\newcommand{\five}{\ts {\sc v}}
\newcommand{\six}{\ts {\sc vi}}
\newcommand{\seven}{\ts {\sc vii}}
\renewcommand\ion[2]{#1\,\,{\sc{\romannumeral #2}}}

\definecolor{maroon}{rgb}{0.760,0.118,0.337}

\usepackage{xcolor}

\newif\ifshowchanges
\showchangestrue

\def\cm{\mbox{\,cm}}
\def\cm3{\mbox{\,cm$^{-3}$}}

\shorttitle{Discovery of SN 2025mkn}
\shortauthors{Lemon et al.}
\suppressAffiliations

\begin{document}

\title{A Natural $\gtrsim$100× Telescope: Discovery of the Strongly Lensed Type II SN 2025mkn at $z = 1.37$}

\correspondingauthor{Ariel Goobar}
\email{ariel@fysik.su.se}

\author[0000-0003-2456-9317]{Cameron~Lemon}
\OKC

\author[0000-0002-4163-4996]{Ariel~Goobar}
\OKC

\author[0000-0001-5975-290X]{Joel~Johansson}
\OKC

\author[0000-0002-8380-6143]{Edvard~Mörtsell}
\OKC

\author[0000-0001-6797-1889]{Steve Schulze} 
\CIERA

\author[0000-0002-8977-1498]{Igor Andreoni}
\UNC

\author[0009-0008-2714-2507]{Aleksandra Bochenek}
\LJMU

\author[0000-0003-1325-6235]{Seán J. Brennan}
\MPIA

\author[0009-0001-0574-2332]{Malte~Busmann}
\LMU
\ORIGINS

\author[0000-0002-8262-2924]{Michael Coughlin}
\Minnesota

\author[0000-0001-8372-997X]{Kaustav~K.~Das}
\Caltech

\author[0000-0002-2376-6979]{Suhail~Dhawan}
\Birmingham

\author[0000-0002-4223-103X]{Christoffer~Fremling}
\CaltechPhys
\CaltechOO

\author[0000-0002-3884-5637]{Anjasha~Gangopadhyay}
\OKCAstro
\author[0000-0003-3270-7644]{Daniel~Gruen}
\LMU
\ORIGINS

\author[0000-0002-9364-5419]{Xander~J.~Hall}
\mcwilliams

\author[0000-0002-9017-3567]{Anna Y. Q. Ho}
\Cornell

\author[0000-0002-5619-4938]{Mansi M. Kasliwal}
\CaltechPhys
\author[0000-0001-8472-1996]{Daniel~A.~Perley}
\LJMU
\author[0000-0002-8121-2560]{Mickael~Rigault}
\Lyon

\author[0000-0001-9915-8147]{Genevieve Schroeder}
\Cornell

\author[0000-0002-3321-1432]{Mathew~Smith}
\Lancaster

\author[0000-0003-1546-6615]{Jesper~Sollerman}
\OKCAstro

\author[0000-0001-8426-5732]{Jean J. Somalwar}
\UCB
\KIPAC

\author[0000-0003-2434-0387]{Robert~Stein}
\UMDAstro
\UMD
\GSFC

\author[0009-0005-6323-0457]{Stephen Thorp}
\OKC
\IoACam
\author[0000-0001-6343-3362]{Alice~Townsend}
\HUBerlin
\author[0000-0003-0733-2916]{Jacob~L.~Wise}
\LJMU

\author[0000-0003-1710-9339]{Lin~Yan}
\CaltechOO

\author[0000-0001-5409-6480]{Nikki~Arendse}
\OKC

\author[0000-0001-8018-5348]{Eric C. Bellm}
\DIRAC
\author[0000-0001-9152-6224]{Tracy X. Chen}
\IPAC
\author{Andrew~Drake}
\Caltech

\author[0000-0002-8532-9395]{Frank J. Masci}
\IPAC

\author[0000-0003-1227-3738]{Josiah Purdum}
\CaltechOO

\author[0000-0001-7062-9726]{Roger Smith}
\CaltechOO


\author[0000-0001-9668-2920]{Jason~T.~Hinkle}
\altaffiliation{NHFP Einstein Fellow}
\UIUC
\SkAI

\author[0000-0002-9204-3256]{T.~Emil~Rivera-Thorsen}
\OKCAstro

\author[0000-0003-4631-1149]{Benjamin~J.~Shappee}
\IfA

\author[0000-0002-2471-8442]{Michael~A.~Tucker}
\altaffiliation{CCAPP Fellow}
\OSU




\author{Jessica~Aguilar}
\LBNL

\author[0000-0001-6098-7247]{Steven~Ahlen}
\BU

\author{Greg~Aldering}
\LBNL

\author{Segev BenZvi}
\ROCH

\author[0000-0001-9712-0006]{Davide~Bianchi}
\UniMi
\INAFBrera

\author{David~Brooks}
\UCL

\author{Todd~Claybaugh}
\LBNL

\author[0000-0002-1769-1640]{Axel~de~la~Macorra}
\UNAMIF

\author[0000-0003-0928-2000]{John~Della~Costa}
\SDSU
\NOIRLab

\author{Arjun~Dey}
\NOIRLab

\author{Peter~Doel}
\UCL

\author{Brenna~Flaugher}
\Fermilab

\author[0000-0002-3033-7312]{Andreu~Font-Ribera}
\ICREA
\IFAEBarcelona

\author[0000-0002-2890-3725]{Jaime~E.~Forero-Romero}
\UniandesPhys
\UniandesObs

\author[0000-0001-9632-0815]{Enrique~Gazta\~naga}
\IEEC
\ICG
\ICE

\author[0000-0003-3142-233X]{Satya~Gontcho~A.~Gontcho}
\UVa

\author{Gaston~Gutierrez}
\Fermilab

\author[0000-0001-6558-0112]{Dragan~Huterer}
\UMichPhys
\UMich

\author[0000-0002-6024-466X]{Mustapha~Ishak}
\UTDallas

\author[0000-0001-8528-3473]{Jorge~Jimenez}
\IFAEBarcelona

\author[0000-0003-0201-5241]{Dick~Joyce}
\NOIRLab

\author[0000-0002-0000-2394]{Stephanie~Juneau}
\NOIRLab

\author{Robert Kehoe}
\SMU

\author[0000-0001-6315-8743]{Alex~G.~Kim}

\author[0000-0002-8828-5463]{David~Kirkby}
\UCIrvine

\author{Theodore Kisner}
\LBNL

\author[0000-0001-6356-7424]{Anthony~Kremin}
\LBNL

\author{Ofer~Lahav}
\UCL

\author[0000-0003-1838-8528]{Martin~Landriau}
\LBNL

\author[0000-0001-7178-8868]{Laurent~Le~Guillou}
\LPNHE

\author[0000-0003-1887-1018]{Michael~E.~Levi}
\LBNL

\author{Marc~Manera}
\IFAEBarcelona
\DeFBarcelona

\author[0000-0002-1125-7384]{Aaron~Meisner}
\NOIRLab

\author{Ramon~Miquel}
\ICREA
\IFAEBarcelona

\author[0000-0002-2733-4559]{John~Moustakas}
\Siena

\author[0000-0001-9070-3102]{Seshadri~Nadathur}
\ICG

\author[0000-0002-9700-0036]{Brendan~O'Connor}
\CMU

\author{Nathalie~Palanque-Delabrouille}
\LBNL
\IRFU

\author{Antonella~Palmese}
\CMU

\author[0000-0002-0644-5727]{Will~J.~Percival}
\WaterlooPhysAstro
\Perimeter
\WaterlooCfA

\author[0000-0001-6979-0125]{Ignasi~P\'erez-R\`afols}
\UPCPhysEEBE

\author{Claire~Poppett}
\LBNL
\UCBSSL
\UCBcampus

\author{Francisco~Prada}
\CSIC

\author{Graziano~Rossi}
\Sejong

\author[0000-0002-9646-8198]{Eusebio~Sanchez}
\CIEMAT

\author{David~Schlegel}
\LBNL

\author{Michael~Schubnell}
\UMichPhys
\UMich

\author[0000-0001-6815-0337]{Arman~Shafieloo}
\KASI
\USTKorea

\author[0000-0002-3461-0320]{Joseph~Silber}
\LBNL

\author{David~Sprayberry}
\NOIRLab

\author[0000-0003-1704-0781]{Gregory~Tarl\'e}
\UMich

\author{Benjamin~A.~Weaver}
\NOIRLab

\author[0000-0002-6684-3997]{Hu~Zou}
\NAOC

\begin{abstract}
We present the discovery of SN\,2025mkn, a gravitationally lensed Type II supernova. First detected as a blue transient in ZTF, 0$\farcs$83 from a $z=0.42$ elliptical galaxy, follow-up SNIFS/UH2.2m and LRIS/Keck spectra revealed absorption lines at $z=1.371$. Later \textit{JWST} NIRCam imaging shows that the bright transient is a close pair of point sources separated by $\sim 0\farcs07$, and a 30 times fainter counterimage opposite the lens, for which NIRSpec reveals strong H$\alpha$ emission also at $z=1.371$. The lightcurves and spectra are consistent with the Type II supernova source being magnified $\gtrsim$ 100 times, with $\sim$250 required to reconcile its luminosity with that of nearby events such as SN\,2023ixf. Lens models are consistent with such high magnifications, and always show that the faint image arrived first (undetected in earlier ZTF imaging), consistent with the later spectral phase of this fainter image. A fourth image is also predicted and possibly detected in the NIRSpec data. Lightcurve-based time-delay measurements are not possible due to the first image being the faintest, however, the resolved NIRSpec spectra offer a future opportunity for time-delay cosmography through supernova phase measurements.

\end{abstract}

\keywords{Supernovae (1668), Gravitational lensing (670), SN2025mkn}

\vspace{2.0cm}
\section{Introduction \label{sec:intro}}
Strong gravitational lenses are unique astrophysical tools: they allow direct probes of total masses within the Einstein radius \citep[e.g.,][]{shajib2022}, reach resolutions impossible for unlensed sources thanks to the associated magnification \citep[e.g., parsec-scales at $z=2$,][]{vanzella2022}, and provide multiple sub-kiloparsec lines of sight to study spatial variations in kinematics and chemistry in galaxies \citep[e.g.,][]{cristiani2024}, among many other uses. Nearly all strong lens systems discovered to date have galaxy or quasar sources \citep{lemon2024}, the latter of which are time variable, opening up additional science cases such as microlensing studies to probe compact matter fractions in lensing galaxies \citep[e.g.,][]{vernardos2024}, and time-delay cosmography, in which the time delays between image pairs are measured and used to constrain the Hubble constant \citep{refsdal1964}. For the latter application \citet{refsdal1964} originally proposed supernova sources, but only in recent years have such systems been discovered \citep{quimby2013, kelly2015, goobar_2017}. These lensed supernovae, despite only providing time-domain information during the lifetime of the supernova, offer several advantages over lensed quasars: (i) time-delay measurements are often easier due to the known shape of the supernova lightcurve, (ii) time delays can be measured independently from spectral phases \citep{bayer2021, johansson2021, 2024ApJ...970..102C}, (iii) some SNe are `standardisable', providing absolute magnifications to break degeneracies of lens modelling \citep{2024MNRAS.531.4349W}, and (iv) after the supernova has faded the lensed host galaxy can be used to constrain the lensing potential without the noise and systematics of overlapping bright point sources \citep[as used for iPTF16geu in, e.g., ][]{mortsell2020}.

While cluster-scale lensed supernovae have already been used for time-delay cosmography \citep[e.g.,][]{vega2018, Kelly_2023, pascale_2024, pierel2025}, the mass models are complex and do not completely use the information of the lensed arcs. Galaxy-scale lensed supernovae offer a promising route to the Hubble constant, thanks to their simple lens mass distributions, though the first examples discovered had time delays too short to offer competitive cosmological constraints \citep{goobar_2017, goobar_2023}. Recently \citet{johansson2025} and \citet{taubenberger2026} reported a lensed Type I superluminous supernova at $z=2.01$ with a maximum image separation of 4.9 arcseconds.




In this letter we report the discovery of SN\,2025mkn at $z=1.371$, a Type II supernova that is multiply imaged by a foreground galaxy at $z=0.42$, with lens coordinates of R.A., Dec. (J2000) = 16:42:11.47, +55:31:02.45. In Section \ref{sec:obs} we outline both ground-based and \textit{JWST} imaging and spectra, and in Section \ref{sec:analysis} we present analysis of the source, the lensing galaxy, and present a simple lens model of the system. We discuss the data, model, and implications in Section \ref{sec:discussion}, and conclude in Section \ref{sec:conclusion}. Throughout this work, we adopt a flat $\Lambda$CDM cosmology with $\Omega_{\mathrm{m}} = 0.3$, $\Omega_{\Lambda} = 0.7$, and $H_{0} = 70\,\mathrm{km\,s^{-1}\,Mpc^{-1}}$. Magnitudes are reported in the AB system.


\section{Observations \label{sec:obs}}
Previous galaxy-scale lensed supernovae have been discovered through an inferred magnification. This requires a redshift and expected absolute magnitude. Ideally, spectra will provide both the redshift and supernova type (and thus expected intrinsic brightness), however, spectra are costly and rare for these types of events. To expand the search, photometric redshifts of potential host galaxies are used so photometric candidates can also be vetted \citep{Goldstein_2017}. In the case of lensing, even though the lensing galaxy will be incorrectly assumed to be a host, it still provides a lower redshift limit with which to search for overly bright supernovae. Our search for lensed supernovae is based on the cadenced survey by the Zwicky Transient Facility (ZTF; \citealt{Bellm+2019,Graham+2019,Dekany+2020,Masci+2019,Patterson+2019,Mahabal+2019,Duev+2019}). 

During our daily automatic cross-match of new ZTF transients to photometric redshifts, SN\,2025mkn -- internally named ZTF25aasjeza -- was alerted as a transient with an intrinsic magnitude of $M=-22.4 \pm 0.1$ given the nearby photometric redshift of $z=0.39\pm0.02$ \citep[Legacy Survey;][]{zhou2021}. A Dark Energy Spectroscopic Instrument (DESI) \citep{DESI2016b.Instr, DESI2022.KP1.Instr, Corrector.Miller.2023, FiberSystem.Poppett.2024, Spectro.Pipeline.Guy.2023, SurveyOps.Schlafly.2023, DESI2024.VII.KP7B, desicollaboration2025datarelease1dark, tr6y-kpc6, Redrock.Bailey.2026} spectrum of the galaxy confirmed it to be $z=0.4203 \pm 0.0001$. The spectrum was taken from the
internal release 'Loa' which will be released as part of DESI DR2 in 2027\footnote{DESI spectrum ID 39633329464544708, available in DESI DR2.}. The transient was also reported by the Asteroid Terrestrial-impact Last Alert System \citep[ATLAS;][]{tonry2018}.

\begin{figure*}
    \centering
    \includegraphics[width=\textwidth]{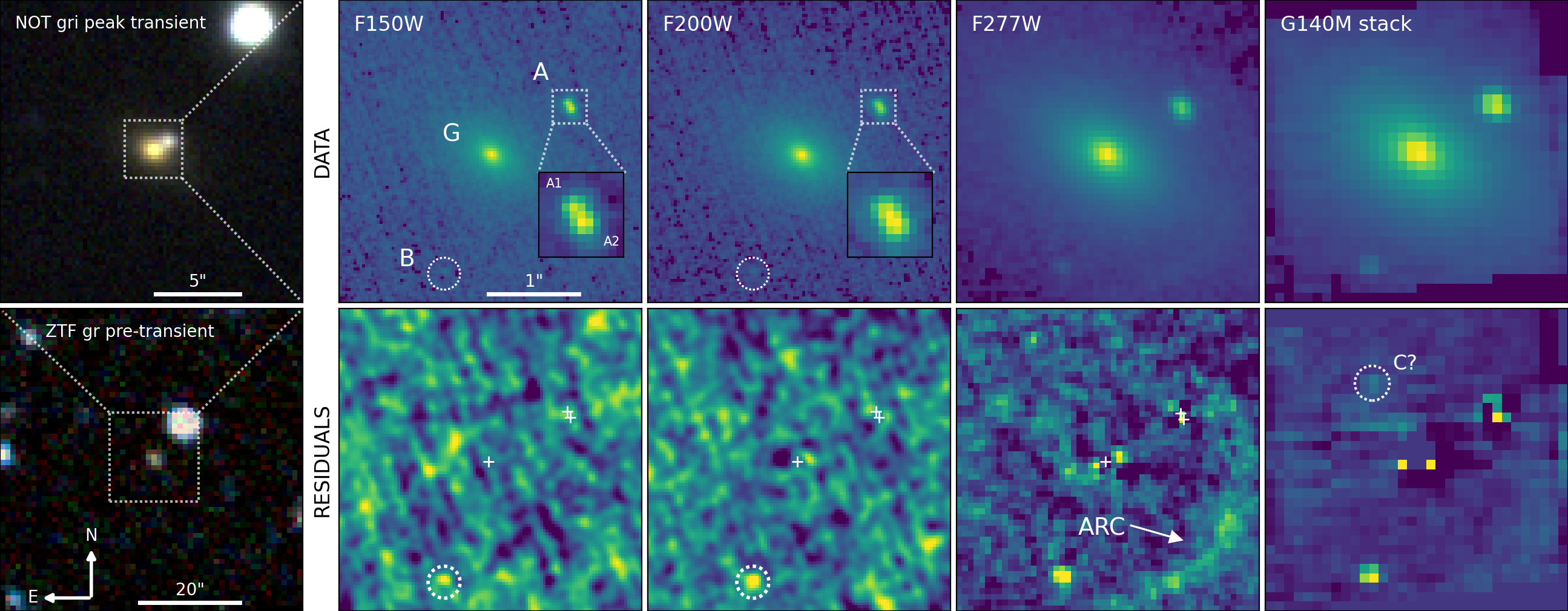}
    \caption{\textit{Left}: ZTF \textit{gr} image of the field before the transient (bottom) and NOT \textit{gri} image during the peak of the lightcurve (top). \textit{Right}: NIRCam F150W, F200W, and F277W images, alongside the NIRSpec G140M white light image (top), and subtractions after modelling the lensing galaxy, G, and image A (bottom). Note the inset showing that image A is well modelled as 2 PSFs separated by $\sim$0\farcs07. The F277W and G140M residuals show an arc to the South West of the galaxy, as well as a hint of a another image, which we label C.
    }
    \label{fig:JWST-data}
\end{figure*}

An early classification spectrum of the transient by the Spectroscopic Classification of Astronomical Transients (SCAT) survey \citep{tucker2022, 2025TNSCR2137....1H} showed a nearly featureless blue spectrum. However, upon closer inspection, narrow absorption lines are present at two distinct redshifts, $z=1.256$ and $z=1.371$. This provides a new lower limit for the supernova redshift of 1.371, implying a rest-frame UV absolute magnitude $M\sim -25$ mag, which is several magnitudes brighter than the most extreme superluminous supernovae; thus, a lensing magnification is required, which was reported in \citet{2025TNSAN.201....1G}. This led to the triggering a \textit{JWST} program, involving imaging which resolves the multiple lensed images (see Figure \ref{fig:JWST-data}) and spectroscopy, both of which we present in this Section, alongside ground-based follow-up. We also obtained Karl G. Jansky Very Large Array (VLA; \citealt{vla, evla}) observations to investigate the source as a possible Luminous Fast Blue Optical Transient (LFBOT), which yielded a non-detection.

The long-wavelength \textit{JWST}/NIRCam data (Figure \ref{fig:JWST-data}) clearly resolve both the lensing galaxy and the transient seen North West of the galaxy. They also show a second point source 1$\farcs$3 South East of the galaxy. We label the two point sources A and B, respectively, and note that B is approximately 30 times fainter than A. We postulate that B is another image of the transient, and provide supporting evidence in the next section.

\subsection{Ground based imaging}
The ZTF \textit{g} and \textit{r} band lightcurves of SN 2025mkn are supplemented by \textit{gri} photometry from both the Spectral Energy Distribution Machine \citep[SEDM,][]{blagorodnova2018, rigault2019, kim2022} on the Palomar 60-inch telescope and the Alhambra Faint Object Spectrograph and Camera (ALFOSC) on the Nordic Optical Telescope, \textit{griz} observations with the IO:O camera on the Liverpool Telescope (LT), and \textit{grzJH} imaging with the Fraunhofer Telescope at Wendelstein Observatory (FTW) using the Three Channel Imager \citep[3KK,][]{Lang-Bardl2016SPIE.9908E..44L}.

For the optical $griz$ data we use archival Pan-STARRS \citep{chambers_pan-starrs1_2016} images for subtractions. For 3KK, the optical CCD and near-infrared (NIR) CMOS data were reduced using a custom pipeline \citep{2002A&A...381.1095G, 2025A&A...701A.225B}. For the astrometric calibration of the images, we used the Gaia EDR3 catalog \citep{Gaia2021, 2021A&A...649A...2L, gaiaEDR3}. Tools from the AstrOmatic software suite \citep{sextractor, scamp, 2002ASPC..281..228B} were used for the coaddition of each epoch's individual exposures (as well as for the IR imaging described later). We use the Saccadic Fast Fourier Transform (\texttt{SFFT}; \citealt{hu_image_2022}) algorithm for image subtraction. 

For the near-IR $J$ and $H-$band data we follow the same image subtraction process as for the optical imaging, but calibrate against the 2MASS Catalog \citep{skrutskie06}. For 3KK, we use templates from the United Kingdom Infrared Telescope \citep[UKIRT;][]{dye_ukirt_2018}.

Due to the high declination of SN 2025mkn, ZTF observations of the field have short seasonal gaps of $\sim$ 55 days. In the discovery season, data starting $\sim$ 170 days before peak are available in the \textit{g} and \textit{r} bands with a median cadence of $\sim$ 2--3 days. In these data we search for any preceding images, and in the case of a non-detection, place upper limits on the brightness.

\begin{figure*}
    \centering
    \includegraphics[width=1.0\textwidth]{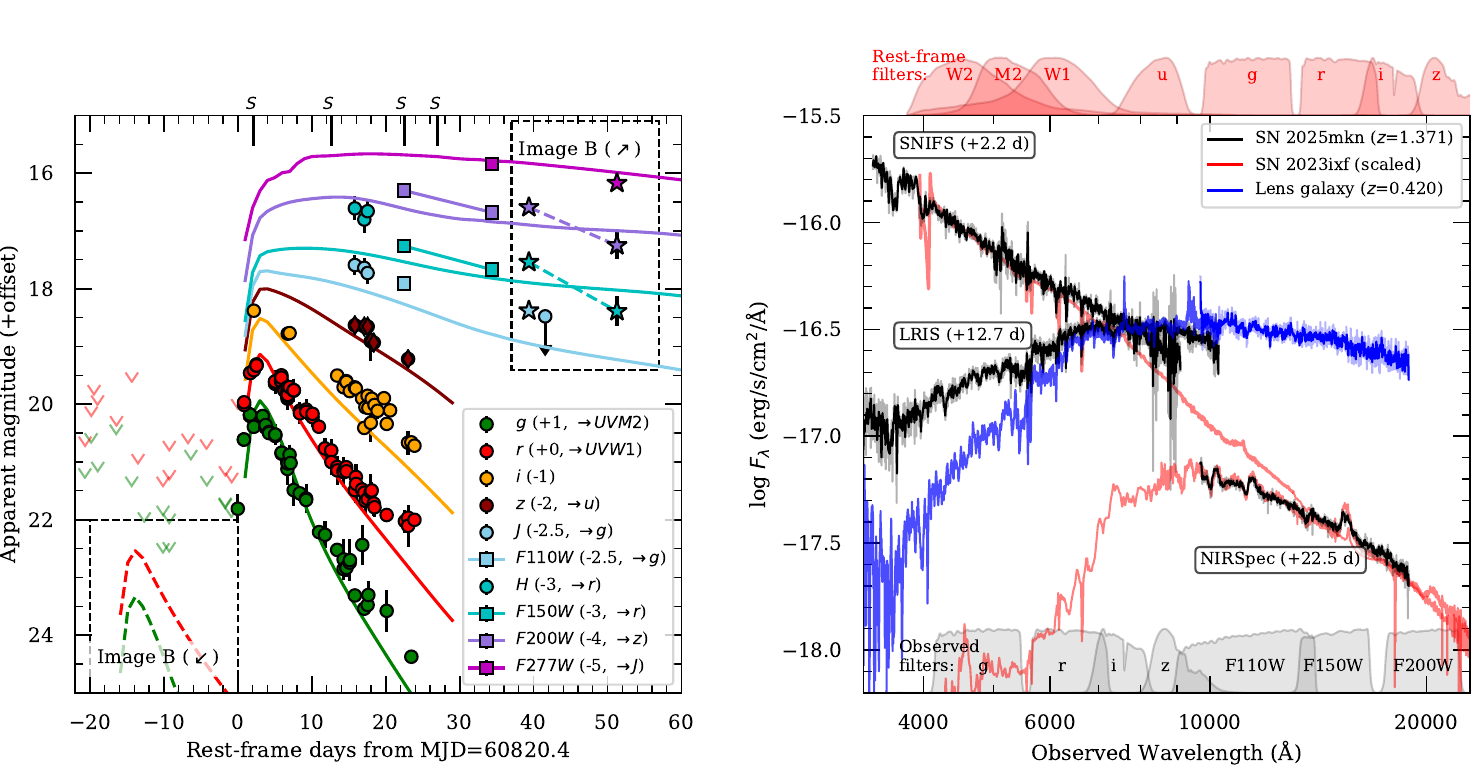} 
    \caption{{\bf Left panel:} Compilation of ground-based (circles) and JWST (square symbols) photometric observations of SN\,2025mkn (Image A). The solid lines show the lightcurves of SN\,2023ixf in matching rest-frame filters or synthetic photometry on spectra indicated in the legend. The star symbols and dashed lines show the photometry for Image B (shifted both in time and magnitude, corresponding to a fiducial time-delay and differential magnification). The observation dates of the spectra are marked along the top with S symbols.
    {\bf Right panel:} Optical and near-IR spectra  (black lines) of SN\,2025mkn (Image A). The shaded red lines show SN\,2023ixf redshifted and scaled to match SN\,2025mkn with magnifications of 250 and 15 (for A and B respectively) at similar phases (reported in rest-frame days).
    The blue line is a combined Keck/LRIS and JWST/NIRSpec spectrum of the lens galaxy.
    }
    \label{fig:lightcurve}
\end{figure*}

We download all epochs of available \textit{g} and \textit{r} imaging from ZTF through the IRSA ZTF API, to build a scene model of the pre-transient system, using \texttt{lightcurver} \citep{dux2024} and \texttt{STARRED} \citep{starred}. After subtracting the best fit constant model (the lensing galaxy and a nearby star), the residuals are inspected for signs of leading images. No detection is seen in any epoch before image A first appears. 5-$\sigma$ non-detections are shown in Figure \ref{fig:lightcurve}. The photometric datasets were collated using $\texttt{fritz}$ \citep{vanderwalt2019, coughlin2023}.

In Figure \ref{fig:lightcurve} we show all the available photometry for this system, with the lightcurves of the nearby Type II supernova, SN 2023ixf, overlaid, with discussion on the nature of the source in Section \ref{sec:source}.

\subsection{Ground based spectra}
On 2025 May 30, a classification spectrum was taken with the SuperNova Integral Field Spectrograph \citep[SNIFS;][]{lantz2004} on the University of Hawaii 2.2 meter telescope through the Spectrosocpic Classification of Astronomical Transients \citep[SCAT;][]{tucker2022}. SNIFS covers the full optical range with both a blue (320 – 560 nm) and red (520 – 1000 nm) channel. The SNIFS spectrum was taken in good conditions, with a seeing of $\approx0\farcs8$, and had an exposure time of 1800 s. The data were reduced with the custom SCAT reduction pipeline, which incorporates wavelength-dependent tracing across each channel and aperture extraction from the reconstructed datacubes \citep[see][for more details]{tucker2022}.

SN2025mkn was later observed with the Low Resolution Imaging Spectrometer (LRIS) on the 10m-class telescope Keck-1 on the nights of 2025 June 24, July 25, and 28, with a $1\farcs0$ wide slit using the B400/3400 grism and R400/8500 grating at parallactic angle. Two exposures of 600 s were taken, and spectra were reduced and extracted using \texttt{LPipe} \citep{Perley2019lpipe}.



\subsection{JWST NIRCam}
\textit{JWST}/NIRCam imaging \citep[programme 3468, ][]{2023jwst.prop.3468G} was taken in simultaneous short- (SW) and long-wavelength (LW) mode on module B with the FULL subarray, using RAPID readout (2 groups per integration; 1 integration per exposure); the SW channel used F150W and F200W, each paired with F277W in LW. We applied an intramodule \texttt{SMALL-GRID-DITHER} (four dithers with sub-pixel steps), obtaining 4 integrations of 21.5 s each, for a total of 86 s in each of the SW filters, and 172 s in the LW filter. We note that the pixel scales for the SW and LW images are 
$0.0312$ and $0.0629\ \arcsec~\mathrm{pix^{-1}}$, respectively. The standard pipeline data showed clear 1/f (flicker) noise \citep{2023zndo...7577320B}, evident as horizontal banding in the SW images, so the pipeline was rerun with \texttt{clean\_flicker\_noise} turned on in stage 1, and the \texttt{skymatch} process in stage 3 \citep{2024PASP..136a5001R}. The data for the system are shown in each filter in the top row of Figure \ref{fig:JWST-data}.

\subsubsection{Light modelling} \label{nircam_light}

As previously mentioned, the F277W data show the bright transient, A, as well as a fainter point source, B, South East of the lens, which we postulate as another image of the source. We also note that A is elongated in the tangential direction, and appears to be consistent with two point spread functions (PSFs) in the F150W and F200W data. We therefore consider these two images -- A1 and A2, as labelled in Figure \ref{fig:JWST-data} -- for the purposes of fitting. To determine galaxy profile parameters and point source positions, we fit three point sources and two concentric Sersic profiles, convolved by the same point source model. We use the \texttt{STPSF DET\_DIST} extension as our PSF model \citep{stpsf}; although this product is formally intended to match detector-coordinate images (e.g. individual rate exposures), we use it here as an approximation for the drizzled PSF. Given the low signal-to-noise of our data, we expect this PSF estimate to be robust for our current analysis.  The galaxy- and A-subtracted residuals (after Gaussian smoothing with widths of 2, 2 and 1 pixels respectively) are shown in the bottom row of Figure \ref{fig:JWST-data}. 

While image B is immediately obvious in the F277W data and the NIRSpec white light image, it is also detected at 5.1- and 5.3-$\sigma$ significance in the $F150W$ and $F200W$ data respectively, even when allowing its position to vary. We report the photometry and astrometry with their 16th and 84th percentile intervals in Table \ref{tab:astrophot}. We also note the presence of an arc-like structure in the South-West of the system only seen in the F277W filter of the imaging data, which we postulate to be the lensed host galaxy of the source supernova.

\begin{table*}
\centering
\begin{tabular}{cccc ccc ccccc}
\toprule
      & \multicolumn{3}{c}{NIRSpec [MJD = 60874]} & \multicolumn{3}{c}{NIRCam [MJD = 60902]} & \multicolumn{5}{c}{Astrometry + Lens model} \\
\midrule
Object & F110W & F150W & F200W & F150W & F200W & F277W & \shortstack{--$\Delta \mathrm{RA}\cos\delta$\\{[arcsec]}} &
\shortstack{$\Delta \mathrm{Dec}$\\{[arcsec]}} & $\kappa$ & $\gamma$ & $\mu$ \\
\midrule
A      & 20.41$^{+0.05}_{-0.05}$ & 20.26$^{+0.05}_{-0.05}$ & 20.30$^{+0.05}_{-0.05}$  & 20.67$^{+0.01}_{-0.01}$ & 20.68$^{+0.01}_{-0.01}$ & 20.84$^{+0.01}_{-0.01}$ & -- & --& -- & -- & --  \\
A1      & -- & -- & --  & 21.33$^{+0.02}_{-0.02}$ & 21.33$^{+0.03}_{-0.03}$ & -- & 0.843$^{+0.003}_{-0.003}$ & 0.474$^{+0.003}_{-0.003}$ & 0.482 & 0.508 & 101 \\
A2      & -- & -- & --  & 21.53$^{+0.02}_{-0.02}$ & 21.54$^{+0.03}_{-0.03}$ & -- & 0.809$^{+0.003}_{-0.003}$ & 0.540$^{+0.003}_{-0.003}$  & 0.492 & 0.518 & $-$97  \\
B      & 24.28$^{+0.05}_{-0.05}$ & 23.94$^{+0.05}_{-0.05}$ & 23.99$^{+0.05}_{-0.05}$  & 24.79$^{+0.26}_{-0.18}$ & 24.65$^{+0.24}_{-0.20}$ & 24.57$^{+0.09}_{-0.07}$ & $-$0.49$^{+0.01}_{-0.01}$ & $-$1.24$^{+0.01}_{-0.01}$  & 0.332 & 0.346 & 3.1  \\
G      & 18.53$^{+0.05}_{-0.05}$ & 18.12$^{+0.05}_{-0.05}$ & 17.90$^{+0.05}_{-0.05}$ & 18.16$^{+0.01}_{-0.01}$ & 17.95$^{+0.01}_{-0.01}$ & 18.00$^{+0.01}_{-0.01}$ &0.000$^{+0.003}_{-0.003}$ & 0.000$^{+0.003}_{-0.003}$  & -- & -- & --   \\
\midrule
A/B flux ratio      & $\sim$35 & $\sim$30 & $\sim$30 & $48^{+13}_{-9}$ & $39^{+6}_{-5}$ & $32^{+3}_{-2}$ & \multicolumn{5}{c}{$68$} \\ 
\bottomrule
\end{tabular}
\caption{Synthetic photometry from NIRSpec (MJD=60873.7) and photometry from NIRCam imaging (MJD=60901.9); for the F110W and F200W synthetic magnitudes are based only on the wavelength range overlapping with the NIRSpec spectrum.  Uncertainties are based on the 16th and 84th percentile intervals for NIRCam and we estimate 0.05 for NIRSpec based on agreement of the galaxy magnitude with NIRCam. The galaxy photometry is based on a 1$\farcs$0-radius aperture.  We also present F200W image astrometry with a fiducial lens model (including convergence, $\kappa$, shear $\gamma$, and image magnification $\mu$), for which a fourth image is also predicted, at $(-0\farcs55, 0\farcs82)$, with $\kappa = 0.632$, $\gamma=0.606$, and $\mu=-4.3$.}
\label{tab:astrophot}
\end{table*}

\subsection{JWST NIRSPEC IFU \label{sec:nirspec}}
Through programme 3468 \citep{2023jwst.prop.3468G}, \textit{JWST}/NIRSpec integral-field spectroscopy was obtained using the G140M grating with the F100LP filter in IFU mode, providing observer-frame spectral coverage from $\lambda \simeq 0.97$ to $1.89~\mu\mathrm{m}$ (R $\sim 1000$). The data were taken in the \texttt{NRS\_FULL\_IFU} aperture as part of a \texttt{4-POINT-DITHER} pattern (a box of $\sim 0\farcs4$ on a side), in NRSRAPID readout with $N_{\mathrm{INT}} = 7$ integrations and $N_{\mathrm{GROUP}} = 12$ groups per integration, resulting in an effective exposure time of $t_{\mathrm{exp}} = 3600~\mathrm{s}$. The data were reduced with the standard \textit{JWST} NIRSpec pipeline \citep{2023zndo...7577320B}, and the spatial pixel scale is $0.10\ \arcsec~\mathrm{spaxel^{-1}}$. 

In many slices, artefacts with diamond-shaped patterns were present -- with the same shape as the dither pattern -- and were masked by hand. For each slice of the combined product with North up, background striping orientated along one axis of the instrument frame is seen. Therefore, in the following modelling process, we fit a varying background perpendicular to this direction, such that at each pixel step along this perpendicular direction, a degree of freedom for a constant background is allowed.

\begin{figure*}
    \centering
\includegraphics[width=1.0\textwidth, trim = 0cm 0cm 0cm 0cm, clip= true]{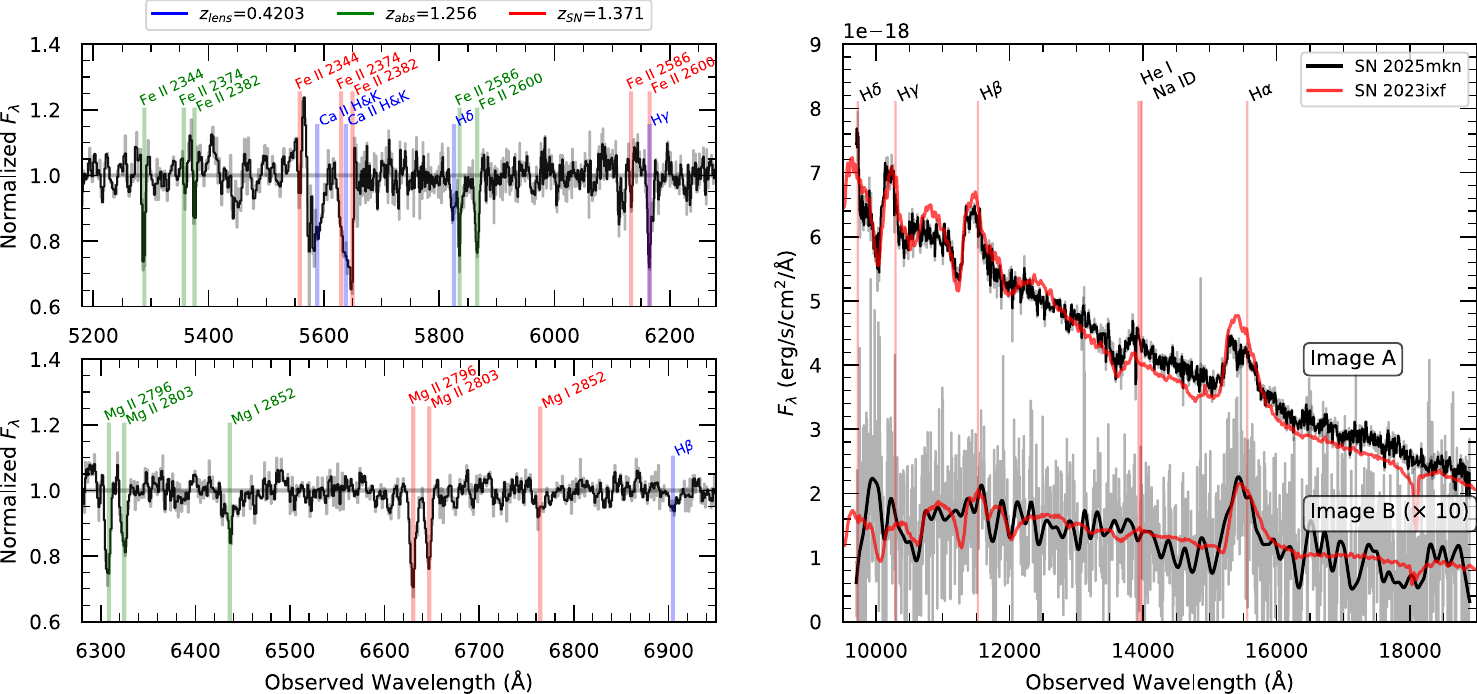}
    \caption{{\bf Left panel:} Selection of narrow absorption features seen in the LRIS spectrum of SN\,2025mkn (here showing the continuum normalized Keck/LRIS spectrum from 2025 June 24). The spectrum shows absorption features from the lens galaxy at $z=0.420$ (blue lines), as well as from systems at $z=1.256$ (green) and $z=1.371$ (red lines). 
    {\bf Right panel:} JWST/NIRSpec spectra of SN\,2025mkn Images A and B (black lines). The red lines show SN\,2023ixf, redshifted to $z=1.371$ and scaled to match SN\,2025mkn. 
    }
    \label{fig:lines}
\end{figure*}

In the white light image of the cube (see Figure \ref{fig:JWST-data}), image B is clearly detected. Image A is unresolved in the NIRSpec image so we extract it with a single PSF, and include another PSF for image B -- as well as a Sersic profile convolved by that same PSF for the lensing galaxy. The PSF is generated using the \texttt{STPSF} package \citep{stpsf}. We use the relative positions as measured from the NIRCAM imaging, only fitting for the galaxy position, so our model at each slice contains two positional degrees of freedom, 5 for the Sersic profile and $\sim$40 for the varying background. The signal of image B is below the noise in most slices, so we repeat the fit by stacking slices in bins of 10 to extract our final spectrum of B. The 2D stacked data (white light) and residuals are shown in Figure \ref{fig:JWST-data}, and we note the presence of the arc which is also seen in the F277W NIRCam data, as well as a possible extra source, which we label as C. The 1D spectrum of A is shown in Figure \ref{fig:lightcurve}, and the spectra of A and B are compared in more detailed in Figure \ref{fig:lines}.

The spectrum of B shows a peak at the wavelength of H$\alpha$ expected at $z\sim1.37$, which now firmly establishes its nature as another image of the source. We therefore also expect that it is fading between the NIRSpec and NIRCam observations (which are separated by 28 days), so we perform synthetic photometry on the NIRSpec 1D spectra of both A and B and present the values in Table \ref{tab:astrophot}. We note that the extraction is reliable as we recover the same value for the F150W magnitude of the lensing galaxy in a $1\farcs0$ circular aperture within 0.05 mag. We take this error as an estimated systematic error on our synthetic photometry. We indeed do see that images A and B fade between the observations, and we discuss this further in Section \ref{sec:discussion}.

\subsection{VLA Observations}

We obtained VLA data of SN2025mkn under Director's Discretionary Time program VLA/25B-366 (PI Ho). The observation started at 2025-09-02 23:29 UT and lasted 2.5 hr at mid-frequencies of 10\,GHz (X-band, 4\,GHz bandwidth) and 15\,GHz (Ku-band, 6\,GHz bandwidth). We calibrated the data using the automated pipeline in Common Astronomy Software Applications package version 6.5.4-9 (CASA; \citealt{casa}) and imaged using standard techniques. We use the \texttt{pwkit/imtool} program \citep{2017ascl.soft04001W} to measure the flux density and image rms. The foreground lensing galaxy is radio bright, and has a 10\,GHz flux density of $205 \pm 60~\mu$Jy and a 15\,GHz flux density of $128 \pm 30~\mu$Jy. There is no apparent emission at the location of the SN2025mkn images, and the rms is measured at $\lesssim 4~\mu$Jy at 10\,GHz and $\lesssim 5~\mu$Jy at 15\,GHz. 

\section{Analysis \label{sec:analysis}} 

\subsection{Spectra and light curves \label{sec:source}}
The early SNIFS spectrum exhibits a steep blue continuum with an inferred blackbody temperature of $\sim$ 27,000 K and an absence of obvious broad supernova features.

Superimposed on the two first spectra are multiple narrow absorption lines. As shown in the left panel of Figure \ref{fig:lines}, we identify prominent \ion{Fe}{2} lines and the \ion{Mg}{2} $\lambda\lambda\,2796,\,2803$ doublet at two distinct redshifts, $z_{1}=1.256 \pm 0.001$ and $z_{2}=1.371\pm0.001$. Additional absorption features at $z_{lens}=0.420$ correspond to the foreground lensing galaxy.

The higher redshift  $z_{2}=1.371$ is adopted as the SN redshift, based on the broad hydrogen features in the \textit{JWST}/NIRSpec spectrum of image A. At a rest-frame phase of +22.5 days, the spectrum displays broad H$\alpha$, H$\beta$ and H$\delta$ emission features, unambiguously identifying SN\,2025mkn as a Type II supernova.
The NIRSpec spectrum of image B likewise exhibits broad $H\alpha$ emission at the same redshift, firmly establishing it as a second lensed image of the same explosion. The presence of fading between the NIRSpec and NIRCam epochs further confirms the transient nature of image B. 


The photometric evolution of image A is shown in Figure \ref{fig:lightcurve}, combining ground-based optical and near-infrared data with resolved \textit{JWST} photometry. The light curves exhibit a close resemblance in shape to those of Type II supernovae, the nearby IIP/IIL SN\,2023ixf in particular \citep{zimmmerman2024Natur.627..759Z}, when compared in matched rest-frame filters. Motivated by this striking similarity, we use SN\,2023ixf as a local analogue to estimate the magnification of SN\,2025mkn. 

Under the assumption that SN\,2025mkn follows the intrinsic luminosity evolution of SN\,2023ixf, the observed peak brightness of image A implies a magnification of $\Delta m_A\sim 6$ mag, corresponding to $\mu_A \sim 250$. Independent estimates from spectral scaling yield consistent results: matching the flux-calibrated spectra of images A and B to SN\,2023ixf at similar phases implies magnifications of $\mu_A \sim 250$ and $\mu_B \sim 15$, respectively. It should be noted that SNe II show significant peak-brightness scatter ($\sigma \sim$ 1 mag), limiting the precision of magnification estimates based on their luminosity.Hence, next we estimate the magnification independent of the assumption that SN 2023ixf and SN 2025mkn are exact analogues in Section \ref{sec:lensmodel} through lens modelling.



The spectral comparison further reveals that image B appears at a later phase than image A, consistent with a significant time delay between the two images. Although the leading image was not detected in archival ZTF data (see inset dashed lines in \autoref{fig:lightcurve}), the relative spectral phases provide an independent handle on the time delay, an approach that will be explored in detail in a forthcoming companion paper (Johansson et al., in prep.).

\subsection{Lens galaxy \label{sec:lensmodel}}

Estimating the total mass and properties of the lensing galaxy is an important step in prioritising follow-up of lensed supernova candidates. In particular, the mass of the lensing galaxy provides an estimate of the image separation of potential multiple images. 

We derive the stellar mass of the system using Pan-STARRS $grizy$ \citep{chambers_pan-starrs1_2016, flewelling20, magnier20}, 2MASS $JHK_S$ \citep{skrutskie06}, and WISE $W1$--$W4$ \citep{wright10} photometry, taken from \texttt{blast} \citep{jones24}. The measured fluxes before extinction correction are ($11.57\pm0.71$, $48.56\pm1.01$, $83.84\pm1.03$, $115.61\pm2.51$, $133.13\pm5.32$, $191.52\pm86.72$, $487.82\pm131.79$, $477.98\pm152.5$, $263.7\pm11.76$, $184.48\pm13.24$, $31.84\pm29.59$, $-103.5\pm258.64$)\,\textmu Jy in ($g$, $r$, $i$, $z$, $y$, $J$, $H$, $K_S$, $W1$, $W2$, $W3$, $W4$). We apply a Milky Way reddening correction using the \citet{gordon23} UV--MIR extinction law, with $E(B-V)=0.012$ mag from the \citet{schlafly11} dust map, with $R_V = 3.1$. We perform a Bayesian SED fit using the \texttt{prospector} modelling framework \citep{johnson21} with a stellar population synthesis (SPS) model based on the Flexible Stellar Population Synthesis code \citep[FSPS;][]{conroy09, conroy10a, conroy10b}, and a parameterisation based on Prospector-$\alpha$ \citep{leja17, leja19}. We use a \citet{Chabrier2003} stellar initial mass function, and a 7-bin star formation history with bins logarithmically spaced in lookback time \citep{leja19_sfh, leja19}. Nebular emission is included based on the \texttt{cloudy} \citep{ferland13} model grid from \citet{byler17}. We carry out posterior sampling using the \texttt{dynesty} \citep{speagle20} nested sampler, with a sampling strategy based on \citet{skilling04, skilling06}, \citet{feroz09}, and \citet{higson19}. We estimate a lens mass of $\textrm{log}_{10} (M/M_{\odot}) \approx 11.4 \pm 0.1$, which provides an estimated Einstein radius of $(1.04 \pm 0.12) \arcsec$
(ignoring aperture corrections and the presence of dark matter).
%

We can independently determine the Einstein radius using the velocity dispersion from the DESI spectrum: 281$\pm$50 \kms. Given the redshifts of the lens and source, and under the assumption of an isothermal mass profile, we constrain $\theta_{E} = (1.38 \pm 0.45)$ \arcsec, in agreement with the estimate from the stellar mass estimate.

Multiple images in strongly lensed systems are typically separated by twice the Einstein radius -- therefore, at least 2$\arcsec$ in the case of SN 2025mkn. Given that image A lies only $\sim0\farcs8$ from the galaxy, another more distant image is expected opposite the lens. Coupled with the magnification argument from the source intrinsic magnitude, our Einstein radius estimate convincingly demonstrated a case of strong lensing, which was borne out by the discovery of image B in the \textit{JWST} datasets.


\subsection{Lens model \label{sec:lensmodel}}

Before we discuss the nature of the system based on the above analysis, we will provide a lens model based on the astrometry of the SW NIRCam imaging.

As previously discussed, image B is a counterimage of the supernova due to the Einstein radius estimates, spectroscopic redshift from NIRSpec, and its transient nature (fading between datasets). We also take A1 and A2 to be separate images of the transient since they are each consistent with point sources, and have the same F200W--F150W colour based on the resolved \textit{JWST} NIRCam imaging.

To elucidate the nature of the system, we fit a simple lens model of a singular isothermal ellipsoid with external shear to recover these three images using \texttt{lenstronomy} \citep{lenstronomy1, lenstronomy2}, but do not place any constraints on the required flux ratios. We only have 6 degrees of freedom (from three image positions), so we fix the lens mass to that measured from the light, i.e., the centroid and flattening parameters -- 150 degrees North of East with an axis ratio of 0.634 -- and only fit for 5 parameters: the Einstein radius, external shear strength and position angle, and a source position. 

\begin{figure}
    \centering
    \includegraphics[width=\linewidth]{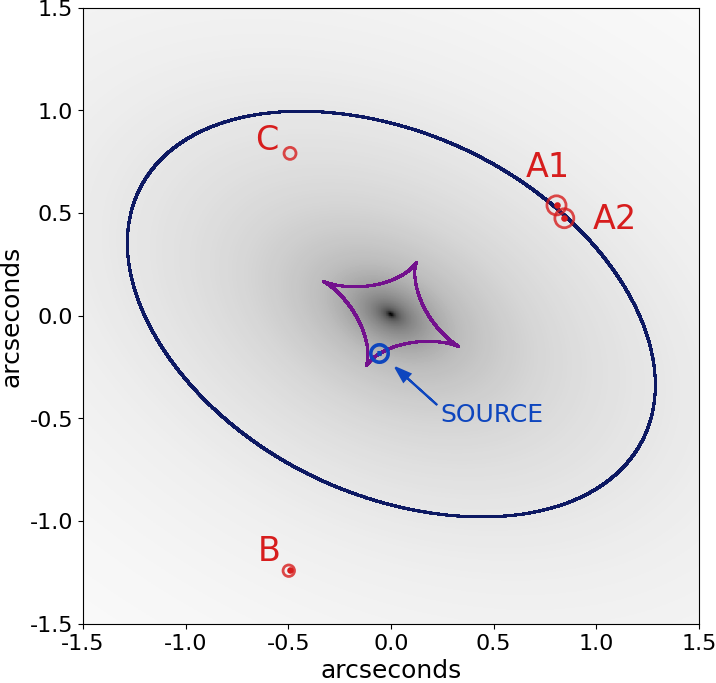}
    \caption{Best-fit elliptical power-law with external shear mass model for SN 2025mkn, with predicted image positions as open red circles. The model recovers the observed positions (red dots) of A1, A2 and B, but also predicts a fourth image, C. The source lies close to the astroid caustic (purple), and the high magnification images of A straddle the critical curve (dark blue).}
    \label{fig:lensmodel}
\end{figure}

We show our best-fit lensing model in Figure \ref{fig:lensmodel}, demonstrating the recovery of the image positions and large magnifications of images A1 and A2. The reduced $\chi^2$ is 2.1. We note that in all models a fourth image is predicted, which we label C, and note that the time delays are similar to those of images A1 and A2, and thus should be closer to peak than image B. Even ignoring this phase, the lens model predicts it to be brighter than image B, however, it is not seen in any of the imaging data. Its predicted position is compatible with a potential detection in the NIRSpec white light residuals (see Figure \ref{fig:JWST-data}), as described in Section \ref{sec:nirspec}. Values for the convergence, shear, and magnifications are provided in Table \ref{tab:astrophot}. The total magnification of our best-fit model is $\sim$200, and the predicted flux ratio of A (A1+A2) to B is $\sim65$. While this is at odds with the observed upper limit of 30 from the NIRSpec flux ratio, we find that much lower total magnifications are possible when the lens mass is allowed to deviate from the light. We discuss alternative explanations for this in Section \ref{sec:discussion}, while keeping the lens mass and light aligned.

The model Einstein radius is $\sim1\farcs12$, and the external shear strength is 0.03 at 128 degrees North of East. This small external shear gives confidence to our assumption that the lens mass follows the light. While our model predicts a time delay of weeks between images A and B, we do not report an exact value here for the sake of blinding any potential time-delay measurements from spectral-phase-matching of the spectra of images A and B. However, we do note that the predicted time delay between images A1 and A2 is approximately 1 minute -- not measurable.

More detailed lens models, which include the information in the lensed host galaxy arc, will be presented in an upcoming paper alongside the spectral time-delay measurements.

\section{Discussion \label{sec:discussion}}

The magnification estimate of $\times$250, based on the assumption of a SN\,2023ixf analogue, is in agreement with the light-mass-aligned lens models, which can predict total magnifications up to 300. However, there are two problems when confronting the lens model to the data: (i) image C is predicted to be brighter than B but is not seen in the imaging data (an upper limit from the F277W data for C is around 0.25 times the brightness of B, or 1.5 magnitudes fainter); and (ii) the flux ratio of A (A1+A2 combined) to B is measured to be around 25 accounting for the time delay, but predicted to be $\sim65$ from the lens model. These discrepancies can be explained by a different lens model in which the lens mass is allowed to deviate from the light. Such models (while under-constrained with just the three image positions) can reproduce the observed A to B ratio of 25--30, though C is still expected at a similar brightness to B. 

We can resolve flux ratio problems in either case by invoking microlensing of a combination of images B and C. Images A1 and A2 are unlikely to be significantly affected by microlensing due to their similar flux ratio -- as predicted by the model. From the NIRSpec white light image we estimate the brightness of the potential detection of image C is around 25\% the brightness of B, consistent with the lack of detection in the NIRCam imaging. We therefore require the B/C brightness to be boosted by a factor of $\sim$4. While reducing the flux of image C can explain this, also boosting the flux of image B can help explain the A/B ratio in the mass-light-aligned models.

We generate microlensing maps given the best-fit $\kappa$-$\gamma$ values from our model \citep{microlensingtool}, assuming a smooth matter fraction of 0.5 (typical for images of similar lenses) and stellar masses of 0.2 $M_{\odot}$ (such that the Einstein radii match the photospheric scale of a type II supernova 40--50 days post-explosion). Image C is demagnified by 0.8 mag or more in 49\% of cases \citep[demagnification is common for saddle points, ][]{schechter2002}, while B is only magnified by 1 mag or more in 6\% of cases. By drawing random realisations of the microlensing in images B and C, and comparing the relative probability of these cases to the most probable case, we find that requiring the above (de-)magnifications is more probable than 20\% of such realisations.

This shows that the observed flux ratios can be satisfactorily explained with microlensing, though the exact reliance on this requires a better-constrained mass model, which will be the focus of future investigations using the information of the extended host arc, and possibly by extracting a position constraint from image C.


\section{Conclusions \label{sec:conclusion}}
We have presented the discovery of SN 2025mkn as a gravitationally lensed Type II supernova, with a magnification of $\times$100 or larger, potentially reaching $\sim$250 if interpreted as a near-perfect analog of SN\, 2023ixf. This system is unique not only as the most magnified supernova yet discovered, but as the first single galaxy lensed supernova with a time delay of weeks. These extreme magnification events are expected in magnitude-limited surveys, such as the previous galaxy-scale spectroscopic discoveries \citep{goobar_2017, goobar_2023}, however, in this case the flux ratio between images is extreme, and might not be expected in mock catalogues that require several, or leading, images to be brighter than a certain threshold \citep[e.g.,][]{om10}.
The striking photometric and spectroscopic resemblance with SN 2023ixf, one of the best studied Type II supernovae in the local Universe, from the rest-frame far-UV to the near-IR, emphasizes the power of lensing magnification to constrain evolutionary effects in stellar explosions over 9 Gyrs. Such detailed comparisons for unlensed supernovae are at the limits of observing capabilities of existing telescopes, as demonstrated by the study of a lensed Type II supernova at $z=5$ \citep{coulter2026}. Triggering follow-up for such systems as soon as possible is key to maximising their scientific use. Robust photometric or spectroscopic redshifts of early type galaxies, and estimates of Einstein radii from stellar mass estimates or velocity dispersions help prioritise the limited spectroscopic follow-up.

One major difficulty of turning this system into a cosmographic probe is the leading image being too faint to obtain a resolved lightcurve. Despite only triggering follow-up long after it faded, spectroscopic phase retrieval offers a way to measuring the time delay, especially in the case of an archival local analogue with well-sampled UV spectra.
In an upcoming paper we will present detailed lens models including modelling of the extended arc, alongside time-delay measurements from the resolved \textit{JWST} NIRSpec spectra, and any associated cosmological constraints.

\normalsize
\vspace{1.5cm}
\section*{Data Availability}
Some of the data presented in this article were obtained from the Mikulski Archive for Space Telescopes (MAST) at the Space Telescope Science Institute. The specific observations analysed can be accessed via \dataset[DOI: 10.17909/xvy2-wa13
]{https://doi.org/10.17909/xvy2-wa13}

\section*{Acknowledgments}

We thank the Space Telescope Science Institute (STScI) support staff for their assistance with the planning and execution of the JWST observations, and for helpful guidance throughout the observing program.

Based on observations obtained with the Samuel Oschin Telescope 48-inch and the 60-inch Telescope at the Palomar Observatory as part of the Zwicky Transient Facility project. ZTF is supported by the National Science Foundation under Award 2407588 and a partnership including Caltech, USA; Caltech/IPAC, USA; University of Maryland, USA; University of California, Berkeley, USA; University of Wisconsin at Milwaukee, USA; Cornell University, USA; Drexel University, USA; University of North Carolina at Chapel Hill, USA; Institute of Science and Technology, Austria; National Central University, Taiwan, and OKC, University of Stockholm, Sweden. Operations are conducted by Caltech's Optical Observatory (COO), Caltech/IPAC, and the University of Washington at Seattle, USA. 

SED Machine is based upon work supported by the National Science Foundation under grant 1106171.

The Liverpool Telescope is operated on the island of La Palma by Liverpool John Moores University in the Spanish Observatorio del Roque de los Muchachos of the Instituto de Astrofisica de Canarias with financial support from the UK Science and Technology Facilities Council.

Based on observations made with the Nordic Optical Telescope, owned in collaboration by the University of Turku and Aarhus University, and operated jointly by Aarhus University, the University of Turku and the University of Oslo, representing Denmark, Finland and Norway, the University of Iceland and Stockholm University at the Observatorio del Roque de los Muchachos, La Palma, Spain, of the Instituto de Astrofisica de Canarias. The NOT data were obtained under program ID P70-501.

Some of the data presented herein were obtained at the W.~M. Keck Observatory, which is operated as a scientific partnership among the California Institute of Technology, the University of California, and NASA. The Observatory was made possible by the generous financial support of the W.~M. Keck Foundation. The authors wish to recognize and acknowledge the very significant cultural role and reverence that the summit of Maunakea has always had within the indigenous Hawaiian community. We are most fortunate to have the opportunity to conduct observations from this mountain.

This material is based upon work supported by the U.S. Department of Energy (DOE), Office of Science, Office of High-Energy Physics, under Contract No. DE–AC02–05CH11231, and by the National Energy Research Scientific Computing Center, a DOE Office of Science User Facility under the same contract. Additional support for DESI was provided by the U.S. National Science Foundation (NSF), Division of Astronomical Sciences under Contract No. AST-0950945 to the NSF’s National Optical-Infrared Astronomy Research Laboratory; the Science and Technology Facilities Council of the United Kingdom; the Gordon and Betty Moore Foundation; the Heising-Simons Foundation; the French Alternative Energies and Atomic Energy Commission (CEA); the National Council of Humanities, Science and Technology of Mexico (CONAHCYT); the Ministry of Science, Innovation and Universities of Spain (MICIU/AEI/10.13039/501100011033), and by the DESI Member Institutions: \url{https://www.desi.lbl.gov/collaborating-institutions}. Any opinions, findings, and conclusions or recommendations expressed in this material are those of the author(s) and do not necessarily reflect the views of the U. S. National Science Foundation, the U. S. Department of Energy, or any of the listed funding agencies.

The Photometric Redshifts for the Legacy Surveys (PRLS) catalog used in this paper was produced thanks to funding from the U.S. Department of Energy Office of Science, Office of High Energy Physics via grant DE-SC0007914.

The authors are honored to be permitted to conduct scientific research on I'oligam Du'ag (Kitt Peak), a mountain with particular significance to the Tohono O’odham Nation.

C.L. acknowledges funding from the European Union’s Horizon Europe research and innovation programme under the Marie Sklodovska-Curie grant agreement No. 101105725. A.G.\ acknowledges financial support from the research project grant “Understanding the Dynamic Universe” funded by the Knut and Alice Wallenberg under Dnr KAW 2018.0067, {\em Vetenskapsr\aa det}, the Swedish Research Council through grants project Dnr 2020-03444, the G.R.E.A.T research environment, Dnr 2016-06012, and the Swedish National Space Agency, Dnr 2023-00226. I.A. is supported by the National Science Foundation award AST 2505775, NASA grant 24-ADAP24-0159, Scialog award SA-LSST-2024-102a, and the Discovery Alliance Catalyst Fellowship Mentors award 2025-62192-CM-19. S.D.\ acknowledges support from  UK Research and Innovation (UKRI) under the UK government’s Horizon Europe funding Guarantee EP/Z000475/1. E.M.\ acknowledges support from the Swedish Research Council under Dnr VR 2024-03927. I.A. is supported by the National Science Foundation award AST 2505775, NASA grant 24-ADAP24-0159, Scialog award SA-LSST-2024-102a, and the Discovery Alliance Catalyst Fellowship Mentors award 2025-62192-CM-19. M.W.C. acknowledges support from the National Science Foundation with grant numbers PHY-2117997, PHY-2308862 and PHY-2409481. M.B. is supported by a Wübben Stiftung Wissenschaft Student Grant. Funded in part by the Deutsche Forschungsgemeinschaft (DFG, German Research Foundation) under Germany's Excellence Strategy – EXC-2094/2 – 390783311. This paper contains data obtained at the Wendelstein Observatory of the Ludwig-Maximilians University Munich. We thank Christoph Ries, Michael Schmid, and Silona Wilke for carrying out the observations. J.T.H. acknowledges support from NASA through the NASA Hubble Fellowship grant HST-HF2-51577.001-A, awarded by STScI. STScI is operated by the Association of Universities for Research in Astronomy, Incorporated, under NASA contract NAS5-26555. TER-T is supported by the Swedish Research Council grant \#2022-04805. S.T. has been supported by funding from the European Research Council (ERC) under the European Union's Horizon 2020 research and innovation programmes (grant agreement no.\ 101018897 CosmicExplorer), and from the research project grant `Understanding the Dynamic Universe' funded by the Knut and Alice Wallenberg Foundation under Dnr KAW 2018.0067. M.A.T. is supported by Program number HST-GO-17429.001-A with funding provided through a grant from the STScI under NASA contract NAS5-26555.

\bibliography{references, Lensbib}




\clearpage
\suppressAffiliationsfalse   
\allauthors         

\end{document}